\documentclass[apjl]{emulateapj}
\usepackage{apjfonts}

  \usepackage{amscd}
  \usepackage{amsmath}
  \usepackage{amssymb}
  \usepackage{verbatim}
  \usepackage{array}
  \usepackage{psfrag}

  \newcommand{\gcc}{\ensuremath \, \mathrm{g} \, \mathrm{cm}^{-3}}  

  \shorttitle{Flame-driven deflagration-to-detonation transitions in Type Ia supernovae?}
  \shortauthors{F.~K.~R{\"o}pke}

\begin{document}
\journalinfo{The Astrophysical Journal, \normalfont{668:1103--1108, 2007 October 20}}
\title{Flame-driven deflagration-to-detonation transitions in Type Ia supernovae?}

\author{F. K. R{\"o}pke\altaffilmark{1}}
\affil{Max-Planck-Institut f\"ur Astrophysik,
              Karl-Schwarzschild-Str. 1, D-85741 Garching, Germany}
\altaffiltext{1}{Also at: Department of Astronomy and Astrophysics, University
  of California Santa Cruz, 1156 High Street, Santa Cruz, CA 95064,
  U.S.A.}
\begin{abstract}
Although delayed detonation models of thermonuclear explosions of white
dwarfs seem promising for reproducing Type Ia supernovae, the
transition of the flame propagation mode from subsonic deflagration to
supersonic detonation remains hypothetical. A potential instant for
this transition to occur is the onset of the distributed burning
regime, i.e.\ the moment when turbulence first affects the internal
flame structure. Some studies of the burning microphysics indicate that
a deflagration-to-detonation transition may be possible here, provided
the turbulent intensities are strong enough. Consequently, the
magnitude of turbulent velocity fluctuations generated by the
deflagration flame is analyzed at the onset of the distributed burning regime in
several three-dimensional simulations of deflagrations in
thermonuclear supernovae. It is shown that the corresponding probability
density functions fall off towards high turbulent velocity
fluctuations much more slowly than a Gaussian distribution. Thus, values claimed to be
necessary for triggering a detonation are likely to be found in
sufficiently large patches of the flame. Although the microphysical
evolution of the burning is not followed and a successful
deflagration-to-detonation transition cannot be guaranteed from
simulations presented here, the results still indicate that such
events may be possible in Type Ia supernova explosions.
\end{abstract}


\maketitle

\setcounter{page}{1103}

\section{Introduction}
\label{intro_sect}

The question of whether deflagration-to-detonation transitions (DDTs) may
occur in type Ia supernova (SN~Ia) explosions has been a puzzle since
the scenario of delayed detonations was suggested in the first place
\citep{khokhlov1991a}. If a supersonic detonation
burning mode occurs at all in SNe~Ia, such a transition is inevitable
because a prompt detonation fails to produce the observed
intermediate mass elements \citep{arnett1969a}. For
these to be synthesized, burning must partially take place at low fuel
densities. 
A pre-expansion of the white dwarf (WD) material before burning is
only possible for subsonic propagation of the thermonuclear
flame. Consequently, the standard picture is that of an accreting WD 
which getting close to the Chandrasekhar limit becomes unstable to a
thermonuclear runaway and ignites a subsonic deflagration flame.
In contrast to the shock-driven detonation, flame
propagation is mediated by microphysical transport in this mode. 
However, this scenario is intricate because the energy released in
burning may expand the WD
so rapidly, that the slow flame is not capable of burning sufficient
amounts of material before it has dropped to densities at which
burning cannot be sustained any longer. This way, the explosion
would be far too weak to be consistent with observations. Therefore,
the flame needs to accelerate and two possible mechanisms
for this have been suggested.

The first is
well-founded on known physical principles and is based on the fact
that the buoyant rise of the burning bubbles generates strong
turbulent motions due to Rayleigh-Taylor and Kelvin-Helmholtz
instabilities. The flame interacts with eddies of the resulting
turbulent cascade and, 
by means of surface area increase, accelerates significantly. This turbulent
deflagration scenario has been recently studied in detail in three-dimensional
simulations \citep[e.g.][]{reinecke2002d,gamezo2003a,roepke2005b}. It
has been shown that it may lead 
to explosions of the WD star meeting the gross properties of
observed SNe~Ia \citep{roepke2007c}. However, some issues remain unsolved in this
model. First, in its current implementations it can explain only the weaker events. Second, it
may in some configurations leave behind unburnt material in the
central parts of the ejecta, which disagrees with observations \citep{kozma2005a}. Third,
it has problems explaining the high-velocity intermediate mass
elements seen in the spectra of SNe~Ia.

These weaknesses of the models may possibly be cured if a
second way of flame acceleration, a DDT, takes place in later stages of the explosion. 
One-dimensional parameterizations of this ``delayed detonation model''
showed best agreement with observations assuming this transition to
occur once the
fuel density ahead of the flame has dropped to $\sim$$10^7 \gcc$.
Unfortunately, such a transition is hypothetical as of yet, since no
convincing mechanism has been identified that would work robustly in a
SN~Ia. 

\citet {niemeyer1997b} pointed out that the only fundamental change in
the flame properties is found at the
transition from the flamelet regime of turbulent combustion to the
distributed regime. In the first, which holds for most parts of the
explosion process, the interaction of the flame with turbulence is
purely kinematic. The flame front as a whole is corrugated by turbulent
eddies, but its internal structure remains unaffected. At a first
glance this seems surprising, since Reynolds numbers $\mathit{Re}(L)$ at scales of
$L = 10^{7} \, \mathrm{cm}$ typical for the situation in SNe~Ia
are of the order
of $10^{14}$. The corresponding
Kolmogorov scale, $\eta \sim L \,[\mathit{Re}(L)]^{-3/4}$, down to
which the turbulent cascade 
extends, is $\sim 10^{-4} \,
\mathrm{cm}$---much smaller than the flame thickness. However, assuming
Kolmogorov scaling, the
velocity fluctuations (denoted as $v'$ in the following) decrease
with the cubic root of the length scale under consideration within the
inertial range of the turbulent cascade. This constitutes
the so-called Gibson scale, at which the laminar flame speed is
comparable to $v'$. Below the Gibson
scale, the flame burns through turbulent eddies faster than they can
deform it, and therefore they do not affect the flame shape
considerably. However, since the laminar burning velocity and the flame
thickness depend on the fuel density \citep{timmes1992a}, the
Gibson scale decreases in the course of the explosion while the flame
structure becomes wider. Therefore, turbulent eddies will
eventually penetrate the internal flame structure and mix heated
material, fuel, and ashes. 

It turns out that this onset of the distributed burning
regime takes place at fuel densities favored for DDTs in
one-dimensional simulations \citep{niemeyer1997b}.
Studying the microphysics of burning in this regime,
\citet{lisewski2000b} concluded that 
triggering a detonation may be possible for sufficiently strong
turbulence. The velocity fluctuations should be close to $10^8
\, \mathrm{cm} \, \mathrm{s}^{-1}$ (corresponding to $\sim$20\% sound
speed), but at this time the available two-dimensional supernova
simulations seemed to exclude such high turbulent velocities. A
re-analysis of this scenario is underway \citep{woosley2007a} and
arrives at similar conclusions. 
If a mechanism providing a deflagration-to-detonation transition
driven by the flame itself existed, 
the parameter deciding on triggering a detonation would be the strength of the
mixing of the flame structure, which is determined by the magnitude of
$v'$. This value can be estimated from three-dimensional
numerical simulations of the deflagration stage of the explosion and
is the focus of the present paper.

\section{Approach}

\begin{table}
\begin{center}
\caption{Parameters of the three-dimensional deflagration SN~Ia
  simulations used for the analysis.
\label{tab:par}}
\setlength{\extrarowheight}{2pt}
\begin{tabular}{llll}
\tableline\tableline
\multicolumn{1}{p{2em}}{name} &
\multicolumn{1}{p{4.5em}}{spatial coverage} &
\multicolumn{1}{p{6.1em}}{resolution (computational grid cells per dimension)} &
\multicolumn{1}{p{5em}}{central density [$10^{9} \, \mathrm{g}\,\mathrm{cm}^{-3}$]}\\
\tableline
Simulation I & full star & 1024 & 2.9 \\
Simulation II & full star & 640 & 2.0 \\
Simulation III & octant & 256 & 2.9 \\
\tableline
\end{tabular}
\end{center}
\end{table}

\begin{figure}
\includegraphics[width = \linewidth]{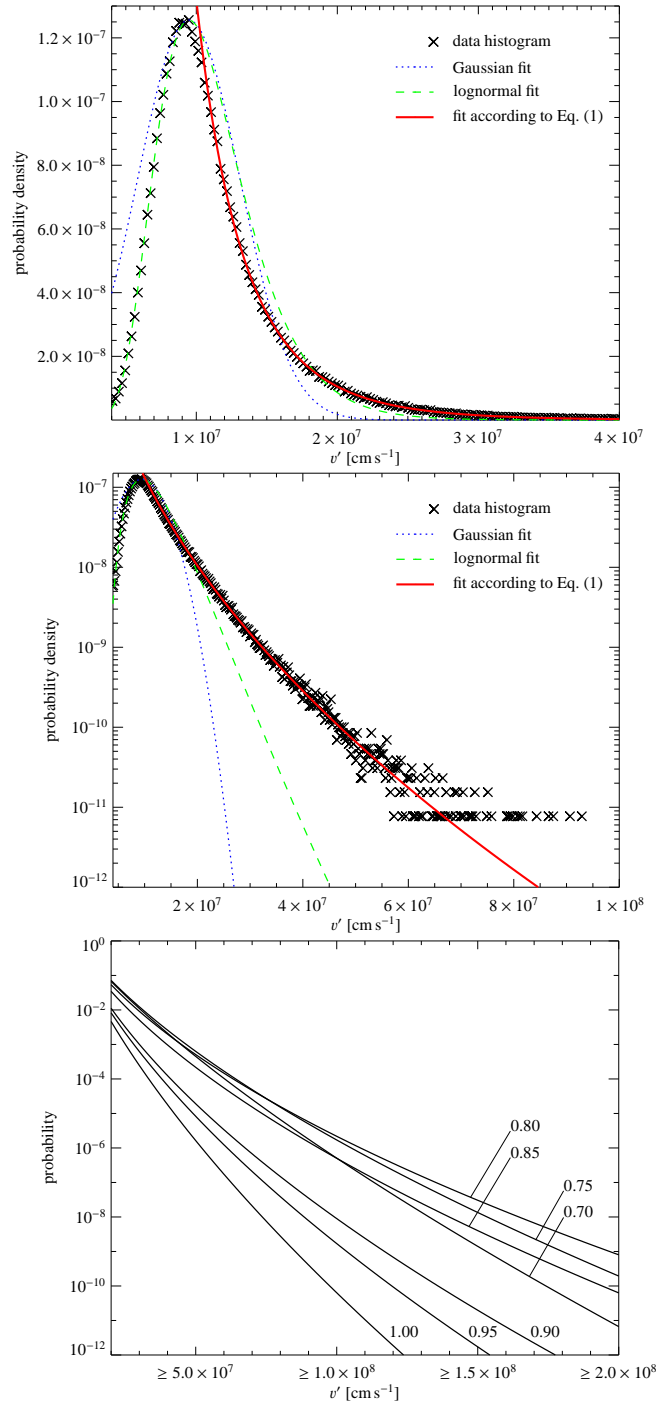}
\caption{Fits to the histogram of $v'$ in simulation I at t = 0.80 s for $1 < \rho
  [10^7\, \mathrm{g}\,\mathrm{cm}^{-3}] \le 3$ \emph{(top two
  panels),} and the
  probability of finding $v'$ larger than a given value according to
  eq.~(\ref{eq:int}) for different
  times in the same density interval \emph{(bottom panel).}\label{fig:fits}}
\end{figure}

\begin{table*}
\begin{center}
\caption{Maximum turbulent velocities at a scale of $10^6 \,
  \mathrm{cm}$, fit parameters according to Eq.~(\ref{eq:fit}), probability of finding $v' \ge 10^8 \,
  \mathrm{cm}\,\mathrm{s}^{-1}$ ($P(10^8)$), estimated flame area
  $A_{\mathrm{est}}$, size of the patch of the flame where  $v' \ge 10^8 \,
  \mathrm{cm}\,\mathrm{s}^{-1}$ ($A_\mathrm{est} P(10^8)$), and number
  of computational cells taken into account in the analysis for
  Simulation I in the density range $1 < \rho
  [10^7\, \mathrm{g}\,\mathrm{cm}^{-3}] \le 3$.
\label{tab:sim1}}
\setlength{\extrarowheight}{2pt}
\begin{tabular}{llllllllr}
\tableline\tableline
\multicolumn{1}{p{2em}}{$t$ [s]} &
\multicolumn{1}{p{4.5em}}{$v'_{\mathrm{max}}(10^6 \, \mathrm{cm})$
  [$10^7 \, \mathrm{cm}\, \mathrm{s}^-1$]} &
\multicolumn{1}{p{5em}}{$-a_0$ [$10^{-4}$]} &
\multicolumn{1}{p{4.67em}}{$a_1$ [$10^{-1}$]} &
\multicolumn{1}{p{4.67em}}{$-a_2$}&
\multicolumn{1}{p{4em}}{$P(10^8)$}&
\multicolumn{1}{p{4.5em}}{$A_\mathrm{est}$ [$\mathrm{cm}^2$]} &
\multicolumn{1}{p{4.5em}}{$A_\mathrm{est} P(10^8)$ [$\mathrm{cm}^2$]}&
\multicolumn{1}{p{3em}}{cells}\\
\tableline
0.70& 6.70 &$     0.356691\pm    0.0004$&$      7.12569\pm    0.0006$&$      12.4931\pm
     0.007$&$ 5.062E-07$&$  7.85E+16$&$  3.98E+10$&       55,341\\
0.75& 8.38 &$      2.97325\pm     0.006$&$      5.98318\pm     0.001$&$      11.1650\pm
      0.02$&$ 1.702E-06$&$  5.35E+17$&$  9.11E+11$&      274,015\\
0.80& 11.5 &$      15.3360\pm     0.010$&$      5.12493\pm    0.0004$&$      9.90905\pm
     0.007$&$ 2.179E-06$&$  1.50E+18$&$  3.28E+12$&      574,769\\
0.85& 10.1 &$      8.92170\pm      0.02$&$      5.44507\pm     0.001$&$      10.3121\pm
      0.02$&$ 5.016E-07$&$  2.52E+18$&$  1.27E+12$&      736,593\\
0.90& 7.05 &$      8.92137\pm     0.006$&$      5.55688\pm    0.0004$&$      9.47029\pm
     0.007$&$ 8.966E-09$&$  2.98E+18$&$  2.67E+10$&      614,195\\
0.95& 6.20 &$      5.18888\pm      0.01$&$      5.87899\pm     0.001$&$      9.71264\pm
      0.02$&$ 1.691E-09$&$  2.89E+18$&$  4.89E+09$&      443,823\\
1.00& 4.24 &$      2.20126\pm     0.005$&$      6.39985\pm     0.001$&$      9.98804\pm
      0.02$&$ 6.351E-11$&$  2.28E+18$&$  1.45E+08$&      263,740\\
\tableline
\end{tabular}
\end{center}
\end{table*}

\begin{table}
\begin{center}
\caption{Selected parameters for different density intervals and
  simulations. The notation is the same as in Table~\ref{tab:sim1}.
\label{tab:all}}
\setlength{\extrarowheight}{2pt}
\begin{tabular}{lllr}
\tableline\tableline
\multicolumn{1}{p{2em}}{$t$ [s]} &
\multicolumn{1}{p{4.5em}}{$v'_{\mathrm{max}}(10^6 \, \mathrm{cm})$
  [$10^7 \, \mathrm{cm}\, \mathrm{s}^{-1}$]} &
\multicolumn{1}{p{7em}}{$A_\mathrm{est} P(10^8)$ [$\mathrm{cm}^2$]}&
\multicolumn{1}{p{3em}}{cells}\\
\tableline
\multicolumn{3}{l}{Simulation I, $1 < \rho
  [10^7\, \mathrm{g}\,\mathrm{cm}^{-3}] \le 2$}\\
\tableline
0.70& 5.37 &$  6.72E+05$&        5723\\
0.75& 7.47 &$  1.05E+11$&       80,438\\
0.80& 11.5 &$  1.66E+12$&      276,683\\
0.85& 10.1 &$  1.20E+12$&      443,841\\
0.90& 6.63 &$  1.82E+10$&      423,654\\
0.95& 6.20 &$  9.74E+09$&      355,975\\
1.00& 4.24 &$  2.58E+08$&      243,480\\
\tableline
\multicolumn{3}{l}{Simulation I, $2 < \rho
  [10^7\, \mathrm{g}\,\mathrm{cm}^{-3}] \le 3$}\\
\tableline
0.70& 6.70 &$  6.00E+10$&       49,618\\
0.75& 8.38 &$  4.40E+11$&      193,577\\
0.80& 9.30 &$  9.83E+11$&      298,086\\
0.85& 7.79 &$  3.06E+11$&      292,752\\
0.90& 7.05 &$  1.78E+09$&      190,541\\
0.95& 4.81 &$  6.96E+08$&       87,848\\
1.00& 3.22 &$  4.92E+07$&       20,260\\
\tableline
\multicolumn{3}{l}{Simulation II, $1 < \rho
  [10^7\, \mathrm{g}\,\mathrm{cm}^{-3}] \le 3$}\\
\tableline
0.75& 5.27 &$  4.52E+10$&        8312\\
0.80& 6.65 &$  1.95E+11$&       45,902\\
0.85& 8.36 &$  8.37E+10$&      121,638\\
0.90& 8.46 &$  1.37E+11$&      200,764\\
0.95& 7.02 &$  1.80E+11$&      239,816\\
1.00& 6.65 &$  6.70E+10$&      230,245\\
1.05& 5.79 &$  2.91E+11$&      187,809\\
1.10& 5.18 &$  3.38E+11$&      139,548\\
1.15& 4.45 &$  1.67E+10$&       90,103\\
\tableline
\multicolumn{3}{l}{Simulation III, $1 < \rho
  [10^7\, \mathrm{g}\,\mathrm{cm}^{-3}] \le 3$}\\
\tableline
0.75& 5.18 &$  5.11E+12$&        9865\\
0.80& 5.67 &$  6.17E+10$&       15,073\\
0.85& 6.51 &$  1.39E+12$&       18,209\\
0.90& 5.75 &$  3.80E+12$&       17,166\\
0.95& 4.04 &$  2.61E+11$&       12,589\\
1.00& 3.00 &$  2.10E+08$&        8098\\
\tableline
\end{tabular}
\end{center}
\end{table}

\begin{figure*}[t]
\includegraphics[width = \linewidth]{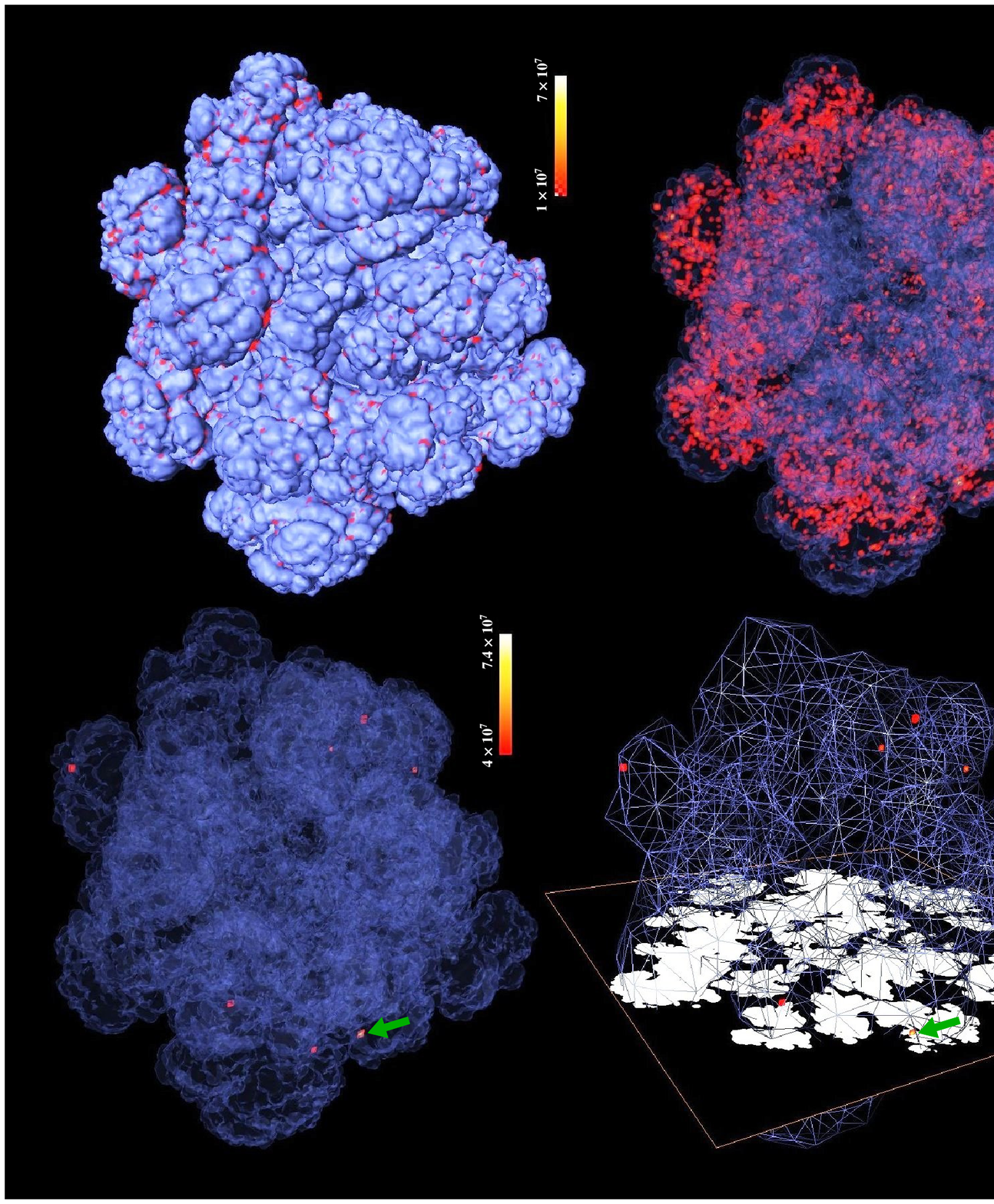}
\caption{Simulation I at $t = 0.80 \, \mathrm{s}$ after ignition. The
  blue opaque, transparent, or wiremesh surfaces correspond to the
  flame front (implemented via a level-set approach). Volumes of high
  turbulent velocity fluctuations are rendered in red/orange. In the
  lower regions, the green arrow indicates the location of the maximum
  value of $v'$ found in the simulation. For better visibility, white
  areas correspond to ash regions and fuel regions are shown in black
  in a plane intersecting with the maximum $v'$-value in the lower
  right region. \label{fig:loc}}
\end{figure*}

To determine $v'$, data
from several three-dimensional simulations of the deflagration stage
in thermonuclear supernovae are analyzed. These simulations apply the methods described in detail by
\citet{reinecke1999a,reinecke2002b}, \citet{roepke2005c}, and \citet{schmidt2006c}. 
The key features include flame tracking via a
level-set approach, a turbulent subgrid-scale model, and a moving
computational grid to follow the expansion of the WD.

Simulation I \citep{roepke2007c} was carried out on $1024^3$ grid cells and
comprised the full star, as did simulation II which was set up on
$640^3$ grid cells. Only an octant of the WD was accounted for in
Simulation III with a $256^3$ cells grid assuming mirror symmetry with
the other octants. The exploding WD had an initial central density of
$2.9\times 10^9 \gcc$ in simulations I and III, while for simulation II a
central density of $2.0 \times 10^9 \gcc$ was chosen. These
parameters of the setup are summarized in Table~\ref{tab:par}.
All initial WD
configurations were composed of a mixture of equal parts, by mass, of
carbon and oxygen. The flame was
ignited in multiple kernels spherically distributed around the center
of the star and partially overlapping, similar to the scenarios
presented by \citet{roepke2006a}.

The quantity under consideration here is $v'$ experienced by the flame in the distributed burning
regime. Of course, this is not directly
resolvable in simulations carried out on the scales of the WD
star. Therefore, an estimate of the turbulent velocities at the flame
is obtained from the subgrid-scale turbulence model. It provides the value of
$v'$ at the scale of the computational
grid. Since the size of the grid cells is a dynamical quantity in the
implementation, it is rescaled to a length scale of $10^6 \,
\mathrm{cm}$ for comparison. To this end, a Kolmogorov
spectrum was assumed, which need not necessarily apply to the actual scaling of
turbulence at the considered spatial range; however, in the
simulations presented by \citet{zingale2005a} and \citet{roepke2007c}.
Kolmogorov turbulence is indeed recovered for buoyancy-unstable flames. In
any case, the
potential error introduced by
this procedure is expected to be small, since the
grid scales at times examined here will not deviate significantly from $10^6 \,
\mathrm{cm}$. Another simplification is introduced by reducing the data sets
by a factor of 2 in each direction, leaving out every second cell of
the computational grid. Although this deteriorates the statistics, it
simplifies the analysis of the large data sets from the highly
resolved simulations.
The distributed burning regime is expected to be reached for fuel
densities in the range $1\ldots 3 \times 10^7 \gcc$
\citep{woosley2007a, niemeyer1997d}. Therefore, we
measure $v'$ determined by the
subgrid-scale model in computational grid cells cut by the flame front
in this density range. This introduces two uncertainties. First, these
cells contain mixed states of fuel and ashes. Consequently, the
values we obtain are not equal (but, due to the small density jump over
the flame, similar to) the values in pure fuel. Although it may seem
compelling to restrict the analysis to cells sufficiently far away
from the flame so that they contain pure fuel, this would require a
distance of at least three cells, and turbulence will be considerably weaker
that far ahead of the flame. Values derived this way would therefore
not well represent the
$v'$ experienced by the flame front. Second, as we are
looking for maxima of $v'$, the
subgrid-scale model will provide only a mean value on the size of the
grid cells. Of course, there will be a maximum of this quantity, which
will be determined in the following, but due to the intermittent
nature of turbulence, rare extrema exceeding these values are possible
on smaller scales in the real situation.

\section{Results and Implications}

The following discussion focuses on simulation I (the
corresponding values are given in Tables~\ref{tab:sim1}
and \ref{tab:all}). Results from the other
simulations are included in Table~\ref{tab:all} and corroborate the
generality of these results. The possibility of triggering a detonation
at a certain turbulence strength depends on the density of
the fuel \citep{lisewski2000b, woosley2007a}.
To account for this density dependence we consider $v'$ at the
flame front in two separate density ranges in simulation I:  $(1 <
\rho_7 \le 2)\, 10^7 \gcc$ and $(2 < \rho_7 \le 3) \, 10^7 \gcc$.

The instantaneous maxima of $v'$ in the snapshots of the simulations
analyzed here fall around $10^8 \, \mathrm{cm} \, \mathrm{s}^{-1}$ 
for all three setups (cf.\ Tables~\ref{tab:sim1}
and \ref{tab:all}). They show a similar temporal evolution passing
through a peak value. 
The maxima of $v'$ are 
slightly lower for less resolved simulations. 
Since the turbulent subgrid-scale model should compensate for
decreasing resolutions, this is not likely due to
lower turbulence prediction. Instead, a lower resolution implies a
less frequent realization of high turbulence
intensities on grid.

\subsection{Probability of high turbulent velocities}

The information gained from determining the maximum turbulent
velocities in the simulation is limited since only stochastic
realizations are recovered.
The robust physical quantity that can be extracted from the simulationss is
the probability density function (pdf) of $v'$. This pdf can be 
approximated with the normalized histogram of  $v'$ determined as described
above. An example is shown in Figure~\ref{fig:fits} \emph{(top)}. 

For determining the possibility of
detonations, the high-velocity tail is the relevant part of the histogram.
An estimate of the corresponding part of the pdf is obtained by fitting
the histogram
with the exponential of a geometric \emph{Ansatz:} 
\begin{equation}
\label{eq:fit}
P(v') = \exp \left[ a_0 (v')^{a_1} +a_2  \right].
\end{equation}
The fit parameters for simulation I at different times are given in
Table~\ref{tab:sim1}.
Integrating this expression from $v$ to $\infty$ yields the
probability of finding $v'$ greater than a
given value $v$,
\begin{equation}
\label{eq:int}
\int_{v}^{\infty} P(v') dv' = \frac{v\, \exp(a_2) \, \Gamma\left( 1/a_1,
    -a_0 v^{a_1} \right)}{a_1 (- a_0)^{1/a_1} v},
\end{equation}
where $\Gamma (.,.)$ denotes the upper incomplete Gamma function.
The results of this fitting procedure for an exemplary data set are
shown in Figure~\ref{fig:fits}. Obviously, Gaussian or lognormal fits
fail to reproduce the high-velocity trend of the histogram (see
Fig.~\ref{fig:fits}, \emph{middle}).

Figure~\ref{fig:fits} \emph{(bottom)} illustrates the time evolution of the
probability of finding a turbulent velocity larger than a given
value, according to equation~(\ref{eq:int}). At earlier times, the
probability of finding $v' \ge 5 \times 10^{7} \, \mathrm{cm} \,
\mathrm{s}^{-1}$ is relatively large, but it decreases steeply for
higher threshold values. Later on, this decrease flattens out
somewhat, but the overall probability becomes lower.

To discuss this behavior, we introduce the probability of finding  $v'
\ge 10^{8} \, \mathrm{cm} \, \mathrm{s}^{-1}$, $P(10^8)$, because
according to \citet{lisewski2000b}, this is approximately the threshold
for triggering a detonation. Values of $P(10^8)$ are given in
Table~\ref{tab:sim1}. For simulation I they follow the temporal trend
of $v'_\mathrm{max}$ and decrease steeply towards late times.
This evolution might be expected. Turbulence decreases in the late
stages of the supernova explosion because the
energy injection from buoyancy instabilities becomes lower and due to
expansion \citep[the explosion ultimately establishes homologous
expansion;][]{roepke2005c}. Thus, the probability of finding high
turbulent velocities 
on any patch of the flame naturally decreases with time.
But here we consider only the part of the flame for which the fuel
densities are in a certain range.
The corresponding flame area increases first and then decreases, since
the fuel densities steadily
decrease in the explosion process. 
This area can be
estimated from the number of computational cells
cut by the flame front and their sizes, providing the value
$A_\mathrm{est}$ given in Tables~\ref{tab:sim1} and \ref{tab:all}.
This temporal evolution of $A_\mathrm{est}$ makes realizations
of a certain high $v'$ more likely at a particular instant given by
the convolution of the area effect with decreasing turbulence
strengths. A quantification is provided in Tables~\ref{tab:sim1} and
\ref{tab:all} with the size of the flame surface $A_\mathrm{est}
P(10^8)$ at which $v' 
>  10^{8} \, \mathrm{cm} \, \mathrm{s}^{-1}$. Obviously,
this area needs to exceed a certain size in order to be relevant for a
DDT. As seen from Table~\ref{tab:sim1}, the evolution of $A_\mathrm{est}
P(10^8)$
follows the maxima of $v'$ determined at different times in
simulation I. However,
for the other simulations 
the trend is less clear. In simulation II the density range of interest is
reached later due to lower initial central density. This delays the
generation of turbulence and thus
burning, energy release, and the expansion of the WD. The overall
lower values of $A_\mathrm{est}
P(10^8)$ derived from simulation II may be taken as an indication that
lower central densities decrease the chances of a DDT, but this
hypothesis needs to be tested in a larger sample of simulations.

\subsection{Location of high turbulent velocities}

In order to establish a self-sustained detonation wave, a minimum mass
of material needs to be burned in the initiation process
\citep{niemeyer1997b,dursi2006a,roepke2007a}. Therefore,
it is necessary that the high turbulent velocities are concentrated in
larger patches and do not distribute all over the flame in tiny
regions.  Although $A_{\mathrm{est}}P(10^8)$ exceeds
$10^{12}\,\mathrm{cm}^2$ in simulations I and III, there is no guarantee that a connected
spatial structure with dimensions larger than
$10\,\mathrm{km}$ exists. 
Moreover, the particular location at the flame
where the high turbulent velocities are found may be critical for the
detonation ignition \citep{woosley2007a}.
Do they occur at leading or trailing edges of
burning bubbles, or in small channels of fuel in between large ash
structures? Such geometrical information cannot be provided by the
pdf.  

To find answers to these questions we consider the situation at $0.8
\, \mathrm{s}$ after deflagration ignition in simulation
I.  

Values of $v'$ exceeding
$10^{8}\, \mathrm{cm}\, \mathrm{s}^{-1}$ are directly found in the simulation
(see Table~\ref{tab:sim1}).
The fact that the grid resolution here is
$\sim$$10^6\,\mathrm{cm}$ indicates that such high turbulent
velocities are localized in finite patches. If the area of high
turbulent velocity fluctuations determined from the pdf was
distributed in many very small patches, it would not be directly
visible in the simulation, due to numerical smoothing. 

Figure~\ref{fig:loc} shows the location of the regions of high
turbulent intensity at the deflagration flame.
The upper left region of Figure~\ref{fig:loc} demonstrates that flame
regions with high turbulent intensities are present
mainly at trailing edges of bubble-like structures. The location of
all regions with turbulent velocities larger than $10^{7} \,
\mathrm{cm} \, \mathrm{s}^{-1}$ is shown again in the upper right
region, where the flame has been made transparent in order to make
these regions better visible. At the instant shown in Figure~\ref{fig:loc}, the
peak value of $v'$ in the analysis of the simulation is found. Its
location is illustrated in the lower regions.

\subsection{Possibility of triggering a detonation}

According to \citet{lisewski2000b}, values of $v'
\gtrsim 10^8 \, \mathrm{cm} \, \mathrm{s}^{-1}$, corresponding to
about 1/5 to 1/4 of the sound speed at the considered densities, are necessary for a
DDT (with optimistic assumptions on the DDT mechanism). The exact
values depend on the fuel density and
composition. For a fuel composition of equal parts, by mass, of carbon
and oxygen (this case applies to the simulations analyzed here), a
threshold of $v' > 5 \times 10^7 \, \mathrm{cm}\,\mathrm{s}^{-1}$ is
given for a fuel density of $2.3 \times 10^7 \, \mathrm{g} \,
\mathrm{cm}^{-3}$. For lower fuel densities, this threshold becomes
larger, and at $8 \times 10^6 \, \mathrm{g} \,
\mathrm{cm}^{-3}$ it is given with $v' > 8 \times 10^7 \,
\mathrm{cm}\,\mathrm{s}^{-1}$. The maxima found in the simulations (in
particular, in simulation I; see Tables~\ref{tab:sim1}
and \ref{tab:all}) exceed these thresholds.

The estimates for the total size of the flame patches where  $v' > 10^8 \,
\mathrm{cm}\,\mathrm{s}^{-1}$ and the fact that high values of $v'$ are found at scales resolved
in the simulation indicate that
the corresponding regions may be sufficiently large to trigger a
detonation (compare with the values of detonator sizes given
by \citet{niemeyer1997b} and \citet{roepke2007a}).

As noted by \citet{lisewski2000b}, the chances of triggering
a detonation increase with higher carbon fractions in the fuel and
decrease with higher oxygen fractions. \citet{roepke2004c}
noted that the energy release in burning does not depend strongly on
the carbon mass fraction of the fuel as long as it proceeds to nuclear
statistical equilibrium. This, however, does not apply to the
situation considered here. The low fuel densities imply incomplete
burning to intermediate-mass
elements. Therefore, in addition to the effect of altering the
thresholds of turbulence strength required for triggering a detonation, 
the carbon mass fraction may also affect the turbulence strength in
the distributed burning regime. Additional studies are required to
settle this question.

\section{Conclusions}
\label{sect:concl}

The intensity of turbulence was analyzed at the onset of the distributed
burning regime in three-dimensional SN~Ia simulations.
Considerably larger values for the maximum
turbulent velocity fluctuations than previously anticipated
\citep{lisewski2000b} were found for all simulations considered and
the conclusion that these values are typical for 
three-dimensional simulations featuring a state-of-the-art treatment
of turbulence on unresolved scales seems compelling. The
histogram of $v'$ features a pronounced high-velocity tail. This part
is not well fit by a Gaussian pdf but rather by an exponential of a
geometric \emph{Ansatz.} 

According to our results, a deflagration-to-detonation transition as
anticipated by \citet{lisewski2000b} is not ruled out. But, lacking a
microphysical model, they cannot provide certainty either, and the
concerns pointed out by \citet{niemeyer1999a} persist.
However, the results presented here encourage detailed studies of the microphysics
of the distributed burning regime at high turbulent velocities, such
as those presented by \citet{lisewski2000b} and \citet{woosley2007a}.

If a DDT occurs in a SN Ia explosion, then a location at the outer
parts of the deflagration flame, but at the trailing edge of a
bubble-like feature, seems most likely. The rapid decline of the pdf
towards high turbulence intensities indicates that such a transition may
be a rare event, possibly realized only once or a few times in a
supernova. This, however, as well as the question of whether a DDT
occurs in every event, depends on the details of the formation of the
detonation and warrants further study.

It may thus be possible that detonations do not form at all leading
features of deflagration flame but only in a few such locations. As a
detonation front cannot cross even tiny regions filled
with nuclear ash \citep{maier2006a}, it may not reach all patches of
unburnt material embedded in the complex deflagration
structure. However, in numerical simulations of
delayed detonations in WD with parametrized DDTs, \citet{roepke2007b}
found that even a 
detonation front formed in a single spot burns most of the remaining
fuel by reaching the dense center of the star.  Such
large-scale supernova simulations with parametrized DDTs
help to address the question of whether
the outcome of delayed detonations is consistent with observational data
\citep{mazzali2007a}.

\begin{acknowledgements}
Stan Woosley greatly supported this work with stimulating
discussions. 
The simulations have been performed on facilities of the Max
Planck Society, Garching, Germany, the HPCx, Edinburgh, UK (as
part of the DEISA project), and at the
National Center for Computational Sciences at Oak Ridge National
Laboratory, which is supported by the Office of Science of the
U.S.\ Department of Energy, under contract  DE-AC05-00OR22725. The
work was supported by the SciDAC Program of the DOE
(DE-FC02-01ER41176) and the 
NASA theory program (NNG 05-GG08G).
\end{acknowledgements}
 




\begin{thebibliography}{25}

\bibitem[{{Arnett}(1969)}]{arnett1969a}
{Arnett}, W.~D. 1969, \apss, 5, 180

\bibitem[{{Dursi} \& {Timmes}(2006)}]{dursi2006a}
{Dursi}, L.~J., \& {Timmes}, F.~X. 2006, \apj, 641, 1071

\bibitem[{{Gamezo} {et~al.}(2003){Gamezo}, {Khokhlov}, {Oran}, {Chtchelkanova},
  \& {Rosenberg}}]{gamezo2003a}
{Gamezo}, V.~N., {Khokhlov}, A.~M., {Oran}, E.~S., {Chtchelkanova}, A.~Y., \&
  {Rosenberg}, R.~O. 2003, Science, 299, 77

\bibitem[{{Khokhlov}(1991)}]{khokhlov1991a}
{Khokhlov}, A.~M. 1991, \aap, 245, 114

\bibitem[{{Kozma} {et~al.}(2005){Kozma}, {Fransson}, {Hillebrandt},
  {Travaglio}, {Sollerman}, {Reinecke}, {R{\"o}pke}, \&
  {Spyromilio}}]{kozma2005a}
{Kozma}, C., {Fransson}, C., {Hillebrandt}, W., {Travaglio}, C., {Sollerman},
  J., {Reinecke}, M., {R{\"o}pke}, F.~K., \& {Spyromilio}, J. 2005, \aap, 437,
  983

\bibitem[{{Lisewski} {et~al.}(2000){Lisewski}, {Hillebrandt}, \&
  {Woosley}}]{lisewski2000b}
{Lisewski}, A.~M., {Hillebrandt}, W., \& {Woosley}, S.~E. 2000, \apj, 538, 831

\bibitem[{{Maier} \& {Niemeyer}(2006)}]{maier2006a}
{Maier}, A., \& {Niemeyer}, J.~C. 2006, \aap, 451, 207

\bibitem[{{Mazzali} {et~al.}(2007){Mazzali}, {R{\"o}pke}, {Benetti}, \&
  {Hillebrandt}}]{mazzali2007a}
{Mazzali}, P.~A., {R{\"o}pke}, F.~K., {Benetti}, S., \& {Hillebrandt}, W. 2007,
  Science, 315, 825

\bibitem[{{Niemeyer}(1999)}]{niemeyer1999a}
{Niemeyer}, J.~C. 1999, \apj, 523, L57

\bibitem[{{Niemeyer} \& {Kerstein}(1997)}]{niemeyer1997d}
{Niemeyer}, J.~C., \& {Kerstein}, A.~R. 1997, New Astronomy, 2, 239

\bibitem[{{Niemeyer} \& {Woosley}(1997)}]{niemeyer1997b}
{Niemeyer}, J.~C., \& {Woosley}, S.~E. 1997, \apj, 475, 740

\bibitem[{{Reinecke} {et~al.}(2002{\natexlab{a}}){Reinecke}, {Hillebrandt}, \&
  {Niemeyer}}]{reinecke2002b}
{Reinecke}, M., {Hillebrandt}, W., \& {Niemeyer}, J.~C. 2002{\natexlab{a}},
  \aap, 386, 936

\bibitem[{{Reinecke} {et~al.}(2002{\natexlab{b}}){Reinecke}, {Hillebrandt}, \&
  {Niemeyer}}]{reinecke2002d}
---. 2002{\natexlab{b}}, \aap, 391, 1167

\bibitem[{{Reinecke} {et~al.}(1999){Reinecke}, {Hillebrandt}, {Niemeyer},
  {Klein}, \& {Gr{\" o}bl}}]{reinecke1999a}
{Reinecke}, M., {Hillebrandt}, W., {Niemeyer}, J.~C., {Klein}, R., \& {Gr{\"
  o}bl}, A. 1999, \aap, 347, 724

\bibitem[{{R{\"o}pke}(2005)}]{roepke2005c}
{R{\"o}pke}, F.~K. 2005, \aap, 432, 969

\bibitem[{{R{\"o}pke} \& {Hillebrandt}(2004)}]{roepke2004c}
{R{\"o}pke}, F.~K., \& {Hillebrandt}, W. 2004, \aap, 420, L1

\bibitem[{{R{\"o}pke} \& {Hillebrandt}(2005)}]{roepke2005b}
---. 2005, \aap, 431, 635

\bibitem[{{R{\"o}pke} {et~al.}(2006){R{\"o}pke}, {Hillebrandt}, {Niemeyer}, \&
  {Woosley}}]{roepke2006a}
{R{\"o}pke}, F.~K., {Hillebrandt}, W., {Niemeyer}, J.~C., \& {Woosley}, S.~E.
  2006, \aap, 448, 1

\bibitem[{{R{\"o}pke} {et~al.}(2007{\natexlab{a}}){R{\"o}pke}, Hillebrandt,
  Schmidt, Niemeyer, Blinnikov, \& Mazzali}]{roepke2007c}
{R{\"o}pke}, F.~K., Hillebrandt, W., Schmidt, W., Niemeyer, J.~C., Blinnikov,
  S.~I., \& Mazzali, P.~A. 2007{\natexlab{a}}, \apj, 668, 1132

\bibitem[{{R{\"o}pke} \& {Niemeyer}(2007)}]{roepke2007b}
{R{\"o}pke}, F.~K., \& {Niemeyer}, J.~C. 2007, \aap, 464, 683

\bibitem[{{R{\"o}pke} {et~al.}(2007{\natexlab{b}}){R{\"o}pke}, {Woosley}, \&
  {Hillebrandt}}]{roepke2007a}
{R{\"o}pke}, F.~K., {Woosley}, S.~E., \& {Hillebrandt}, W. 2007{\natexlab{b}},
  \apj, 660, 1344

\bibitem[{{Schmidt} {et~al.}(2006){Schmidt}, {Niemeyer}, {Hillebrandt}, \&
  {R{\"o}pke}}]{schmidt2006c}
{Schmidt}, W., {Niemeyer}, J.~C., {Hillebrandt}, W., \& {R{\"o}pke}, F.~K.
  2006, \aap, 450, 283

\bibitem[{{Timmes} \& {Woosley}(1992)}]{timmes1992a}
{Timmes}, F.~X., \& {Woosley}, S.~E. 1992, \apj, 396

\bibitem[{{Woosley}(2007)}]{woosley2007a}
{Woosley}, S.~E. 2007, \apj, 668, 1109

\bibitem[{{Zingale} {et~al.}(2005){Zingale}, {Woosley}, {Rendleman}, {Day}, \&
  {Bell}}]{zingale2005a}
{Zingale}, M., {Woosley}, S.~E., {Rendleman}, C.~A., {Day}, M.~S., \& {Bell},
  J.~B. 2005, \apj, 632, 1021

\end{thebibliography}
\end{document}